\documentstyle[12pt]{article}
\def\beq{\begin{equation}}
\def\eeq{\end{equation}}
\def\beqa{\begin{eqnarray}}
\def\eeqa{\end{eqnarray}}

\begin{document}

%%%%%%%%% %%%%%%%%% %%%%%%%%% %%%%%%%%% %%%%%%%%% %%%%%%%%% %%
\title{\sc Testing  the Meson Cloud Model in Inclusive Meson Production} 
\author { F. Carvalho \thanks{e-mail: babi@if.usp.br}, 
\ F.O. Dur\~aes \thanks{e-mail: fduraes@if.usp.br}, 
\ F.S. Navarra \thanks{e-mail: navarra@if.usp.br}, 
\ M. Nielsen \thanks{e-mail: mnielsen@if.usp.br}  \\
{\it Instituto de F\'{\i}sica, Universidade de S\~{a}o Paulo}\\
{\it C.P. 66318,  05315-970 S\~{a}o Paulo, SP, Brazil}} 
\maketitle
\vspace{1cm}
\begin{abstract}
We have applied the Meson Cloud Model  to calculate inclusive
momentum spectra of pions and  kaons produced in
high energy proton-proton and proton-nucleus collisions. For the first time 
these data are used to constrain the cloud  cut-off parameters.  
We show that it is 
possible to obtain a reasonable description of data,
especially the large $x_F$ ($x_F \geq 0.2 $) part of the spectrum and at the same time 
describe (partially) the E866 data on $\overline d - \overline u$ and $\overline d / \overline u$.  
We also discuss the 
relative strength of the $\pi N$ and $\pi \Delta$ vertices. We find out that the 
corresponding cut-off parameters should be both soft and should not differ by more than $200$ $MeV$
from each other.  
An additional source (other than the meson cloud) of 
sea antiquark asymmetry,  seems to be necessary to completely explain
the data. A first extension of the MCM to proton nucleus collisions is discussed. 
\\
PACS numbers 14.20.Dh~~12.40.-y~~14.65.-q
\\

\end{abstract}

\vspace{1cm}
%\newpage

\section{Introduction}

Recent measurements \cite{hawker,peng} have established the asymmetry in the distributions of up
and down quarks in the nucleon sea, a result which can not be understood in terms 
of perturbative QCD. The presence of pions in the nucleon can naturally account for
the excess of $\overline d$ over $\overline u$. The role of mesons in 
deep inelastic scattering (DIS) was first investigated by
Sullivan \cite{sul}. He suggested that 
some fraction of the nucleon's anti-quark sea distribution may be
associated with the pion cloud of
the nucleon. Several works developed this idea and
gave origin to the Meson Cloud Model (MCM) \cite{ku,cloud,nos}. They are all based on the 
notion that the physical proton (p) may be expanded in a sum of virtual meson-baryon 
($MB$) states. The probabilities of these states are not known apriori. They are commonly
related to the probability of the splitting $p \rightarrow M B$, which, in turn, is 
calculated with a simple Feynman diagramm of meson emission. In these calculations the 
assumption is made that the proton is already an extended object and a form factor is 
assigned to the meson emission vertex. This form factor contains a cut-off parameter which
must be adjusted by fitting experimental data. In some works it is adjusted to correctly
reproduce the Gottfried Sum Rule (GSR) violation. The most consistent procedure, however, is 
to fix the cut-off by simultaneously analyzing data on  hadronic collisions and parton distribution 
functions. This is the atitude adopted, for example, in \cite{hss}. 
The cut-off choices vary over a wide range according to the 
experimental source used in the determination procedure. In ref. \cite{koepf} a detailed 
discussion of the subject is presented.

In a very recent analysis of the E866 data \cite{peng}, the authors used the MCM to fit the 
$\overline d(x) - \overline u(x)$ as a function of $x$. The dominant intermediate $MB$ 
states are $\pi N$ and $\pi \Delta$. The conclusion was that the cut-off associated 
with both states must be soft and also $\Lambda_{\pi N} \geq \Lambda_{\pi \Delta}$. In 
particular, with dipole form factor $\Lambda_{\pi N}= 1.0 $ $GeV$ and 
$\Lambda_{\pi \Delta}=0.8$ $GeV$. A 
similar conclusion was reached in \cite{koepf}. With 
such choices it was possible to find ``nearly exact accord'' with experiment. This 
finding on one hand increases the confidence in the virtual meson-baryon picture but
on the other hand imposes constraints on the MCM description of other high energy 
collision data. In many previous studies   
$\Lambda_{\pi N}$ and  $\Lambda_{\pi \Delta}$ were assumed to be roughly equal \cite{hss,ing}. Shortly after the E866 analysis, Melnitchouk, Speth  and Thomas \cite{mt} presented a new
calculation of the $ \overline d - \overline u$ asymmetry in the framework of the MCM, 
including $\pi N$ and $\pi \Delta$ states  and also  Pauli exclusion principle effects. 
Their conclusion was that data can be reproduced with a dipole form factor if  
$\Lambda_{\pi N}= 1.0$ $GeV$  and  $\Lambda_{\pi \Delta}= 1.3$ $GeV$.

In this work we address inclusive meson production at high energy proton-proton and 
proton-nucleus collisions. We concentrate on low $p_T$  ($ p_T \leq 0.3  \,GeV$) and large 
$x_F$ ($ x_F \geq 0.2 $) where non-perturbative effects are dominant and the meson cloud 
is most relevant. More specifically we address the $x_F$
distributions of $\pi^+$ and  $K^+$ produced in $p p$
and $p A$ collisions at $p_{lab} \simeq 100  \,GeV/c$. We present simultaneously a fit of the
meson inclusive spectra and the analysis of the GSR including the recently measured 
$\overline d (x) - \overline u (x)$ distribution. Our purpose is to make a new test of the 
MCM and at the same time check the cut-off choices made in refs. \cite{peng} and 
\cite{mt}.

\section{Meson spectra in the MCM}

Assuming that in proton-proton collisions the proton behaves like a 
meson ($M$)  -  baryon ($B$)  
state, the possible reaction mechanisms for meson production 
at large $x_F$ and small $p_T$ are illustrated in Fig. 1. In Fig. 1a  the baryon  
just ``flies through'', whereas the corresponding meson
interacts inelastically producing a $\pi^+$ or a $K^+$ in the final state.
In Fig. 1b  the meson  just ``flies through'', whereas the corresponding baryon
interacts inelastically producing a $\pi^+$ or a $K^+$ in the final state. In Fig. 1c 
the meson in the cloud is already a $\pi^+$ or a $K^+$ which escapes. We shall refer to 
the  first two processes as  ``indirect production'' and to the last one as ``direct 
production''. The first two are calculated with  convolution formulas whereas the 
last one is given basically by the meson momentum distribution in the cloud  
initial $ | MB > $
state. Direct production has been widely used in the context of the MCM and applied to study
$n$, $\Delta^{++}$ and $\pi^0$ production \cite{hss,nsss,dal}. Indirect meson production has been
considered previously in a simplified approach \cite{sea}.

In ref. \cite{hss} a process analogous to 1c, in which the baryon escapes was used 
to determine various cut-off parameters. The 
authors found  values around $1 \,GeV$  for all of them. However the agreement 
between theory and experiment was  poor. This suggests that the no-rescattering 
assumption is not appropriate. In \cite{dal} the same type of process was considered but 
rescattering of the baryon with the target was included and absorptive corrections were
calculated. This results in a reduction of about $40 \%$ of the cross section at almost all 
values of the baryon fractional momentum ($z$) except at $ z \simeq 1$ where absorptive
corrections are not important. The indirect 
mechanism has not been applied to baryon production because the sub-process 
$ B p \rightarrow B X $ is not experimentally well known. On the other hand we can study
inclusive meson production with 1a) since data on $ M p \rightarrow M X$ are available.   

As stated above, in the MCM the projectile proton is regarded as being a sum of virtual  
meson-baryon pairs and a proton-proton reaction can thus be viewed as
reaction between the ``constituent'' mesons and baryons  of one 
proton  with the other proton.

We shall decompose the proton in the following possible Fock states:
\beqa
|p> &=&  Z \, [ \,\, |p_0> +\,|p_0 \pi^0> +\, |n \pi^+> +\, |\Delta^0 \pi^+> +\, |\Delta^+ \pi^0> 
+\, |\Delta^{++} \pi^->  \nonumber\\  
& & +\, |\Lambda K^+> +\,  |\Sigma^0 K^+> +\,  |\Sigma^{0 *} K^+> +\, 
 |\Sigma^+ K^0> +\,  |\Sigma^{+ *} K^0> \,\, ] 
\label{fock}
\eeqa
where $|p_0>$ is the bare proton. We consider only light mesons. In \cite{hss} the 
state $ |N \rho >$ was also included but was found to be relevant only for production of 
large $p_T$ particles. Since we will be restricted to low $p_T$ production we shall 
neglect the havier mesons.  

The relative normalization of these states 
is in principle fixed once the cloud parameters are fixed. $Z$ is a
normalization constant.

In the $ | MB > $ state  the meson and baryon 
have  fractional momentum $y_M$ and $y_B$ with distributions called  
$f_{M/MB}(y_M)$ and $f_{B/MB}(y_B)$ respectively. Of course $y_M + y_B = 1$ and 
these distributions are related by:
\beq
f_{M/MB}(z) =  f_{B/MB}(1-z) 
\label{fmb}
\eeq

The ``splitting function'' $f_{M/MB} (y)$ represents the probability density to find 
a meson with momentum fraction $y$ of the nucleon and is usually given by 
\begin{equation}
f_{M/MB} (y) = \frac{g^2_{M B p}}{16 \pi^2} \, y \, 
\int_{-\infty}^{t_{max}}
dt \, \frac{[-t+(M_B-M_{p})^2]}{[t-m_{M}^2]^2}\,
F_{M B p}^2 (t)\; ,
\label{fpin}
\end{equation}
for baryons $B$ belonging to the octet and  
\begin{equation}
f_{M/MB} (y) = \frac{g^2_{M B p}}{16 \pi^2} \, y \, 
\int_{-\infty}^{t_{max}}
dt \, \frac{[(M_B + M_{p})^2 - t]^2 [(M_{p} - M_B)^2 - t]}
{ 12 M_B^2 M_{p}^2 [t-m_{M}^2]^2}\,
F_{M B p}^2 (t)\; 
\label{fpidel}
\end{equation}
for baryons belonging to the decuplet.
In the above equations  $t$ and $m_{M}$ are the four momentum square and the mass of
the meson in the cloud state, 
$t_{max}$ is the maximum $t$ given by
\begin{equation}
t_{max} = M^2_B y- \frac{M^2_{p} y}{1-y} 
\label{tmax1}
\end{equation}
$M_B$ ($M_{p}$) is the mass of the baryon $B$ ($p$). Since the function 
$f_{M/MB} (y)$ has the interpretation of ``flux'' of mesons $M$ inside 
the proton, the corresponding integral
\beq
n_{M/MB} = \int_{0}^{1} \,\, d y \,\,f_{M/MB} (y)  
\label{n_m}
\eeq
can be interpreted as the ``number of mesons'' in the proton or ``number of mesons in the air''. In many works the magnitude of the multiplicities $n_{M/MB}$ has been considered a
measure of the validity of MCM in the standard formulation with $MB$ states. If these 
multiplicities turn out to be large ( $\simeq 1$) then there is no justification for
employing a one-meson truncation of the Fock expansion. The model has no longer 
convergence. This may happen for large cut-off values.

The  invariant  cross section for production of 
positively charged mesons $M^+$ ($\pi^+$ or $K^+$) is given by:
\beq
E \frac{ d^3 \sigma^{pp \rightarrow M^+X}}{d p^3} =  \frac{x_F}{\pi}
 \frac{ d \sigma^{pp \rightarrow M^+X}}{d x_F d p^2_T}\,=\,  \Phi_M\,\, +\,\, \Phi_B\,\, 
+\,\, \Phi_D
\label{sechoque}
\eeq
where
\beq 
\Phi_M = \sum_{MB}\int_{x_F}^{1}\, \frac{dy}{y}  f_{M/MB} (y) \, 
\frac{x_F}{\pi y}
\frac{ d \sigma^{M + p \rightarrow M^+X}} {d (x_F/y) d p^2_T} (x_F/y)
\label{dsigmeson} 
\eeq
and
\beq
\Phi_B =  \sum_{MB}\int_{x_F}^{1}\, \frac{dy}{y}  f_{B/MB} (y) \, 
\frac{x_F}{\pi y}
\frac{ d \sigma^{B + p \rightarrow M^+X}} {d (x_F/y) d p^2_T} (x_F/y) 
\label{dsigbaryon}
\eeq
$\Phi_M$ and $\Phi_B$ refer respectively to the indirect meson and baryon 
initiated reactions and 
$x_F$ and $p_T$ are respectively the fractional longitudinal momentum and 
transverse momentum of the outgoing meson. The sum is over all the cloud states
in (\ref{fock}).  

$\Phi_D$ represents the direct process depicted in Fig. 1c and is given by:
\beq
\Phi_D = \sum_{MB} f_{M/MB} (x_F , p_T^2) \,\,\,\, \sigma^{B p} (s_X)\,\, K_{abs}
\label{direct}
\eeq
where $f_{M/MB} (x_F , p_T^2)$ is given by (\ref{fpin}) and (\ref{fpidel}) 
, not integrated over $t$ and with the 
replacement $ t= - p^2_T / (1-x_F) -  x_F^2 M_B^2 / (1-x_F)$. The quantity 
$\sigma^{B p} (s_X) $ is 
the baryon-proton cross section 
at cms energy $\sqrt{s_X}$ and $K_{abs}$ is an absorption factor. This cross section is 
energy ($s_X$) dependent  but in the energy range considered here its variation is very 
small and therefore we take it as a constant  $\sigma^{B p} = 38$ $mb$. 
The $K$ factor was used in refs. \cite{dal,nsz} to account for rescattering of the escaping 
cloud element (in that case a baryon and here a meson) against the proton target. 
It varies in the range $0 \leq K_{abs} \leq 1$ and expresses
the efficiency of the direct process. Of course this is a model dependent quantity. As stated 
above, the modest agreement of direct production calculations with data obtained in several 
works and the improvement obtained in \cite{dal,nsz}  (with the inclusion of absorption effects) strongly suggest that this factor is important and we shall keep it. 

Once the `` splitting functions '' (\ref{fpin}) and (\ref{fpidel})  are known we can 
calculate the part of the
antiquark distribution in the proton coming from the pion cloud with the convolution: 
\beq
\overline q (x)_{f} = \int_{x}^{1} \frac{d y}{y} f_{M/MB} (y)\,\, 
\overline q^{\pi}_{f} (\frac{x}{y})
\label{quark}
\eeq
where $\overline q^{\pi}_{f} (z)$ is the flavour $f$ valence antiquark distribution in the pion.  
With the above formula we can compute the $\overline d$ and $\overline u$ distributions, their  difference, $ D = \overline d(x) - \overline u(x) $, and calculate the Gottfried 
integral:
\beq
S_G = \frac{1}{3} - \frac{2}{3} \int_{0}^{1}  [ \overline d (x) - \overline u(x) ] dx 
\label{gottfried}
\eeq

\section{Inputs for the calculation}

\subsection{``Elementary cross sections''} 

The invariant cross sections for the reactions $M p \rightarrow M^+ X$ 
and $B p \rightarrow M^+ X$, appearing in the convolutions 
(\ref{dsigmeson}) and (\ref{dsigbaryon}) are listed in Table I.  Those in the first 
column can be taken directly from experimental data\cite{bar,bren}. The unmeasured  
$\pi^0 p \rightarrow M^+ X$ process was approximated by the 
average between $\pi^+ p \rightarrow M^+ X$ and $\pi^- p \rightarrow M^+ X$. The same 
procedure was used for the process $ K^0 p \rightarrow M^+ X$.

In making use of experimental cross sections  we are treating the virtual mesons in the 
cloud as real particles. In our case 
this can be justified because the observed mesons both on the left hand side of 
(\ref{sechoque}) and 
on the right hand side of (\ref{dsigmeson}) and of (\ref{dsigbaryon}) have very small 
tranverse momentum. 
Consequently the involved cloud particles must also have small tranverse momentum, being 
therefore not far from the mass shell.  

The reactions in the second and third columns of Table I are not measured. We expect them all
to be of the same order of magnitude (aditive quark model approximation). We therefore 
approximate all these cross sections by an ``average cross section ''.  In the case of $\pi^+$ 
and $K^+$ production we shall assume respectively that:
\beq
\frac{ x_F d \sigma^{Bp \rightarrow \pi^+X}}{\pi d x_F d p_{T}^2}\,\,
\simeq\,\, 
\frac{ x_F  d \sigma^{pp \rightarrow \pi^+X}}{\pi d x_F d p_{T}^2}
\,\,\,\, ; \,\,\,\,
\frac{ x_F d \sigma^{Bp \rightarrow K^+X}}{\pi d x_F d p_{T}^2}\,\,
\simeq\,\, 
\frac{ x_F d \sigma^{pp \rightarrow K^+X}}{\pi d x_F d p_{T}^2}
\label{rep1}
\eeq
This implies that all baryons are equally efficient as the proton to produce $\pi^+$  or $K^+$ in collisions with a target proton.  The absorption factor appearing in 
(\ref{direct}) is chosen to be $K_{abs} = 0.4 $ for pion production and 
$K_{abs} = 0.8$ for kaon production. These values are within the range of theoretical 
estimates presented in \cite{nsz}.

\begin{center}
\begin{tabular}{|c|c|c|}  \hline
$ M p \longrightarrow M^+ X$ & $B p \longrightarrow M^+ X$ & $B p \longrightarrow M^+ X$ \\
\hline
$\pi^+ p \rightarrow M^+ X$ & $ n p \rightarrow M^+ X$ & $ \Lambda p \rightarrow M^+ X$  \\
\hline
$\pi^0 p \rightarrow M^+ X$ & $ p_0 p \rightarrow M^+ X$ & $ \Sigma^0 p \rightarrow M^+ X$ \\
\hline
$\pi^- p \rightarrow M^+ X$ & $ \Delta^0 p \rightarrow M^+ X$ & $ \Sigma^{0*} p \rightarrow M^+ X$ \\
\hline
$ K^+ p \rightarrow M^+ X$ & $ \Delta^+ p \rightarrow M^+ X$  &  $ \Sigma^+ p \rightarrow M^+ X$ \\
\hline
$ K^0 p \rightarrow M^+ X$ & $ \Delta^{++} p \rightarrow M^+ X$  &  $ \Sigma^{+*} p \rightarrow M^+ X$\\
\hline
\hline
\end{tabular}
\end{center}
{\bf Table I:} Subprocesses contributing to inclusive meson production 
($M^+ = \pi^+$ or $K^+$).\\

\subsection{Coupling constants}

The coupling constants   are either measured, inferred from isospin symmetry or 
estimated with, for example, QCD sum rules. We will take them from other 
works \cite{koepf,ing,holtz,hss} and keep them fixed. They are given in Table II.

\begin{center}
\begin{tabular}{|c|c|}  \hline
$g_{p \pi^+ n}$  & $ \sqrt{2} g_{p \pi^0 p} =  \sqrt{2} \, (-3.795 \sqrt{4\pi}) $ \\
\hline
$g_{p \pi^0 p}$  & $  -3.795 \sqrt{4\pi}  $ \\
\hline
$g_{p \pi^+ \Delta^0}$  & $\frac{1}{\sqrt{6}}\,g_{p \pi \Delta} = \frac{1}{\sqrt{6}}\,28.6$ \\
\hline
$g_{p \pi^0 \Delta^+}$  & $\sqrt{2} \,g_{p \pi^+ \Delta^0} = \frac{1}{\sqrt{3}}\,28.6$ \\
\hline
$g_{p \pi^- \Delta^{++}}$  & $\sqrt{3} \,g_{p \pi^+ \Delta^0} = \frac{1}{\sqrt{2}}\,28.6$ \\
\hline
$g_{K^+ p \Lambda} $ & $  -3.944 \sqrt{4\pi} $ \\
\hline
$g_{K^+ p \Sigma^{*0}}$ & $ \frac{1}{\sqrt{3}}\, g_{K p \Sigma^*} =
                 \frac{1}{\sqrt{3}}\, \frac{1}{2} \, g_{p \pi \Delta} = 
                 \frac{1}{\sqrt{3}}\, \frac{1}{2} \, 28.6 $ \\ 
\hline
$g_{K^+ p \Sigma^{0}}$ & $  \frac{1}{\sqrt{3}}\, g_{K p \Sigma} = 
 \frac{1}{\sqrt{3}} \, \frac{\sqrt{3}}{5} \,  g_{p \pi^0 p} = 2.69 $ \\
\hline
$g_{K^0 p \Sigma^{*+}}$ & $ \sqrt{2} \, g_{K^+ p \Sigma^{0*}} =
          \sqrt{2} \,  \frac{1}{\sqrt{3}}\, \frac{1}{2} \, 28.6 $ \\ 
\hline
$g_{K^0 p \Sigma^{+}}$ & $  \sqrt{2}\, g_{K^+ p \Sigma^0} = 
 \sqrt{2} \,\,\, 2.69 $ \\
\hline
\hline
\end{tabular}
\end{center}
{\bf Table II:} Coupling constants.\\

\subsection{Cut-off parameters}

In the calculations we need the baryon-meson-baryon form factors 
appearing in the splitting functions. Following a phenomenological
approach, we use the dipole form:
\begin{equation}
F_{M B p} (t) = \left(  \frac{ \Lambda^2_{M B p} - m_{M}^2} 
                       {\Lambda^2_{M B p} - t} \right)^2
\label{eq:form}
\end{equation}
In the above formula $\Lambda_{M B p}$ is the  form factor cut-off parameter.

In \cite{koepf} an extensive discussion 
concerning the appropriate value of the cut-off has been made, using exponential
form factors. The conclusion of the authors was that, for exponential form factors,
$\Lambda^e_{\pi N N} \simeq 1000 \,MeV$, $\Lambda^e_{\pi N \Delta} \simeq 800 \,MeV$  and 
$\Lambda^e_{K N Y} \simeq 1200 \,MeV$. They find out that allowing the $\pi N N$ and 
$\pi N \Delta$ vertices to be different improves the quality of their fits. On the 
strange sector, on the other hand, they use a harder and universal cut-off. In ref. 
\cite{hss} the MCM was used to study baryon ($n$, $\Lambda$ and $\Delta^{++}$) 
spectra in $p p$ collisions and the resulting fits strongly suggest a universal cut off 
$\Lambda^e \simeq 1000 \,MeV$.  
The exponential form factor is, of course, not the only possible choice. One might 
use a monopole or  dipole form as well. Differences between the
various forms are not particularly important and, besides, as it was
pointed out by Kumano \cite{ku2} it is possible to ``translate''
one form factor with its cut-off to another form factor with a
correspondingly different cut-off, the overall results being 
approximately equivalent. The approximate relation between the exponential ($e$), 
dipole ($d$) and monopole ($m$) cut-offs is given by $
\Lambda^m \simeq 0.62 \Lambda^d  \simeq  0.78 \Lambda^e
$. As pointed out in \cite{koepf}, in dealing
with decuplet splitting functions monopole form factors lead to divergencies. With 
an exponential form factor one avoids this problem but the calculations can only be
done numerically. Using the dipole form factor we can perform the integrations 
(\ref{fpin}) and (\ref{fpidel}) analytically. Because of this advantage we choose 
the dipole form.

Once translated to the monopole form the cut-off values quoted above give values 
around $800 \,MeV$  which are still significantly smaller than old analyses 
performed with the Bonn potential \cite{holtz} or the Nijmegen \cite{nim}
potential which favour a harder  cut-off  ($\Lambda^m \simeq 1000 \,MeV$) but are
close to values obtained in more recent analyses performed by the same group 
\cite{jan}.

In view of all the works done so far on this subject, we may conclude that 
these cut-off parameters must be soft. The next question which is now under 
debate is: which cut-off is larger, $\Lambda_{\pi N N}$  or $\Lambda_{\pi N \Delta}$ ?
As pointed out in the introduction, the E866 analysis \cite{peng} favours 
$\Lambda_{\pi N N} \geq \Lambda_{\pi N \Delta}$ whereas Melnitchouk, Speth and Thomas 
\cite{mt} suggest that $\Lambda_{\pi N N} \leq \Lambda_{\pi N \Delta}$. 

Inspired on these two works we shall test the following choices for the dipole 
cut-off parameters:
\beqa 
(I)\,\, \Lambda_{oct} = 0.96 \,\,  GeV  \,\,\,\, \Lambda_{dec} = 0.77 \,\, GeV  \nonumber \\
(II)\,\, \Lambda_{oct} = 0.87 \,\, GeV \,\,\,\,  \Lambda_{dec} = 1.0 \,\, GeV
\label{lambchoice}
\eeqa
$\Lambda_{oct}$ and 
$\Lambda_{dec}$ are the cut-off parameters for all the octet and decuplet vertices 
respectively. As it will be seen, 
these  choices are in reasonable agreement with experimentally measured meson $x_F$ 
spectra. 

\section{Results and discussion}

Before presenting a comparison with experimental data we show in Fig. 2 the behaviour 
of the splitting functions $f_{M/MB}$ and $f_{B/MB}$ with respect to cut-off variations. 
Solid and dashed lines represent respectively the octet eq.(\ref{fpin}) and decuplet
eq.(\ref{fpidel}) $MB$ states. The two upper (lower) boxes show functions calculated 
with soft (hard) cut-off parameters. The upper and lower boxes on the left (right) 
show functions with  octet cut-off parameters harder (softer) than the decuplet ones.
The values, chosen for the purpose of illustration, are:  
$\Lambda_{oct} = 1.0 $ $GeV$ and $\Lambda_{dec} = 0.80 $ $GeV$ (Fig. 2a);  
$\Lambda_{oct} = 0.80 $ $GeV$ and $\Lambda_{dec} = 1.0 $ $GeV$ (Fig. 2b);  
$\Lambda_{oct} = 2.0 $ $GeV$ and $\Lambda_{dec} = 1.6 $ $GeV$ (Fig. 2c);  
$\Lambda_{oct} = 1.6 $ $GeV$ and $\Lambda_{dec} = 2.0 $ $GeV$ (Fig. 2d). We consider the
momentum ($y$) distributions of $\pi^+$ in the states $\pi^+ n$ and $\pi^+ \Delta^0$ and 
also the baryon momentum ($y_B=1-y$) distribution in these states. Because of   
eq.(\ref{fmb})  all these figures are symmetric and the meson momentum distributions 
(always on the left part of each box) are the ``mirror'' pictures of the corresponding 
baryon momentum distributions (always on the right part of each box).  
The comparison between the upper pannels with the lower ones shows that at soft  
cut-off's there is a more distinct separation between baryons and mesons, the former
peaking at higher values of $y$ and the latter at smaller $y$'s.  At harder cut-off's 
all the curves tend to become identical and centered at $y = 0.5$.  The decuplet 
functions lie always bellow the octet functions, except if 
$\Lambda_{dec}  \gg \Lambda_{oct}$, which is unrealistic. In Fig. 2d we can see that
$f_{\pi^+ / \pi^+ \Delta^0} (y)$ lies above $f_{\pi^+ / \pi^+ n}$ for $y \geq 0.6$.  
This feature was 
found in \cite{mt} to be very important in the study of sea antiquark asymmetry. 

Using, as input, the splitting functions shown in Figs. 2a, 2b, 2c and 2d in 
eq. (\ref{sechoque}) for $M^+ = \pi^+$ we obtain the pion spectra shown in Figs. 
3a, 3b, 3c and 3d respectively. They are decomposed in meson initiated ($M$), 
eq.(\ref{dsigmeson})(dot-dashed lines), baryon initiated ($B$), eq.(\ref{dsigbaryon})
(dashed lines), direct ($D$), eq.(\ref{direct})(dotted lines), and the total sum ($T$). 

We see that the direct process dominates pion
production up to $x_F \simeq 0.7$. In the two lower figures (harder cut-off parameters) 
they are so important for all $x_F$ that one can neglect the $M$ and $B$ contributions. For 
soft cut-off choices the $B$ component becomes important too and even dominant at 
$x_F \geq 0.8$ (shaded area in Figs. 3a and 3b). Comparing the two shaded areas, we 
observe that with   $\Lambda_{\pi N N} \leq \Lambda_{\pi N \Delta}$  (on the right) 
the $B$ component starts  earlier to dominate the pion $x_F$ spectrum. This component is 
the harder and flatter one and is essential to obtain a good description of data at 
large $x_F$. The direct contribution alone would fall too fast and underestimate data. 
Neglecting the direct contribution leads to a good fit only with the $M$ and $B$ components 
but at the expense of unrealist large (dipole) cut off choices:  
$\Lambda_{oct} = 1.13 $ $GeV$ and $\Lambda_{dec} = 1.8$ $GeV$. Only with these large numbers 
we recover the normalization of the experimental spectrum.  Neglecting also the $B$ component
leads to  a good fit with the $M$ component alone and a cut-off $ \Lambda_{oct} = 2.1 $ $GeV$.
In short, it is clear that we
need all three components to describe data, the first one ($D$) to get the proper normalization
and the last two ($B$ and $M$) to get the correct shape of the meson spectra.  In this way we get 
also reasonable pion (and also kaon) multiplicities. For  both cases (I and II) we obtain 
$n_{\pi} \simeq 0.17$, which is small enough to justify the use of the one- meson truncation of
the Fock decomposition.

In Fig. 4a and 4b we present our pion and kaon spectra compared with experimental data, 
circles from ref. \cite{bar} and triangles from ref. \cite{bren}. In these figures we
include all three components and plot eq.(\ref{sechoque}) with the cut-off choices 
I (solid line) and II (dashed line). As it can be seen, the overall 
agreement with data is good over a large range on $x_F$. We have checked that significant 
deviations from these choices lead to large disagreement between our spectra and data. As an
example we show in Fig. 4 with dotted lines the spectra corresponding to the 
the choice $\Lambda_{oct} = 1.0$ $GeV$ and $\Lambda_{dec} = 1.3$ $GeV$ 
discussed in \cite{mt}. As it can be seen they are already somewhat far from 
data points, especially in the case of the kaon spectrum. From this figure 
we may conclude that the MCM can give 
an accurate description of data with soft ($\Lambda \leq 1.0$ $GeV$) cut-off parameters. Choosing 
$\Lambda_{oct} \geq \Lambda_{dec}$ or
$\Lambda_{oct} \leq \Lambda_{dec}$ does not make much difference as long as both 
cut-off's differ only by $\simeq 200$ $MeV$. 

In Figs. 5a and 5b we show the difference $\overline d - \overline u$ and ratio 
$\overline d / \overline u$ respectively. The quark distribution functions were calculated 
with the convolution formula (\ref{quark}). The points represent the E866 data. Solid and
dashed lines represent the  parameter choices I and II respectively. As expected our solid 
lines are very close to the ones presented by the E866 analysis in \cite{peng}. Using a 
larger cut-off for the $\pi N \Delta$ vertices leads to the dashed curves.  As pointed out
in \cite{mt} this choice  leads to a reduction in the antiquark asymmetry, leaving room for
additional sources of asymmetry, such as the Pauli exclusion principle. 

The value of the Gottfried integral given by  (\ref{gottfried}) is 
$S_G = 0.261  $ (case I) and $S_G = 0.314 $ (case II).  The value reported by the NMC collaboration is
$ 0.235 \pm 0.026 $. Taken together, Figs. 4 and 5 and the integral $S_G$, it seems that choice I
is better everywhere but fails badly in the $ \overline d / \overline u$ ratio. Choice II, on the 
other hand, is never  far from data points but gives a too large value for $S_G$. 
We thus conclude that
a better global description of data can be obtained with 
$\Lambda_{oct}$ moderately smaller than $\Lambda_{dec}$ and that some additional source of asymmetry, other than the meson cloud,  is  required. 

Another interesting test for the MCM is inclusive meson production in proton 
nucleus collisions. In principle we can use the same formulas eqs. (\ref{sechoque}), 
(\ref{dsigmeson}), (\ref{dsigbaryon}) and (\ref{direct}), replacing the cross sections  
$\sigma^{ M (B) + p \rightarrow M^{+} X }$ by $\sigma^{ M (B) + A \rightarrow M^{+} X }$ and 
keeping exactly the same splitting functions. However most of the 
$ M (B) + A \rightarrow M^{+} X $ cross sections  are not measured. 
As a first test we shall assume, following 
the experimental analysis described in refs. \cite{bar,bren}, that:
\beq 
E \frac{ d^3 \sigma^{pA \rightarrow M^+X}}{d p^3} \simeq \,\, const \,\,  A^{\alpha}  
\,\, E \frac{ d^3 \sigma^{pp \rightarrow M^+X}}{d p^3} 
\label{sechoca}
\eeq
where the constants are chosen to reproduce the normalization of data points.

Usually $\alpha$ is a very slowly varying function of $x_F$. We take it constant 
($\alpha = 0.72$). The main goal of this exploratory study is only to test
the shape of the spectra obtained with the MCM. 
In the above expression the $p p$ cross sections are exactly the same as before.  
In Fig. 6 we show the spectra obtained with (\ref{sechoca}) for pions (6a and 6b) and 
for kaons (6c and 6d).  The cut-off choice I (II) is depicted with solid (dashed) line. Since 
the $|MB>$ state will interact inside the nucleus it may be affected by 
medium effects. The simplest way to incorporate these effects is by changing the baryon masses 
according to $M_B \rightarrow M^*_B \simeq 0.8 M_B$. With this modification the meson splitting 
functions will increase in normalization and will peak at larger values of $y_M$. Because of this
the final meson spectra will also peak at larger values of $x_F$. The inclusion of medium effects    
(curves in dotted lines in Fig. 6) improves the agreement between the MCM and data.

\section{Conclusions}

This work was motivated by the recent revival of the meson cloud 
picture of the nucleon. We performed a comparison between inclusive meson spectra 
predicted by the MCM and experimental data. Instead of using only the  ``direct process'',  
as done in refs. \cite{dal,hss} we have included the ``indirect process'', in which
the final cross section is a convolution of the MCM ``splitting functions'' with the
``elementary'' meson-nucleon and baryon-nucleon cross sections, in an 
analogous way as done in the QCD 
parton model calculations. In this way we can test the universal splitting functions 
and determine the cut-off parameters. This determination is done by simultaneouly 
analysing mesonic spectra, difference and ratio of antiquark distributions and the Gottfried 
sum rule.

The first conclusion of this work is that the MCM 
describes reasonably  well the production of pions and kaons in $p p$  high
energy collisions. Even the intermediate $x_F$ region ($ 0.2 \leq x_F \leq 0.6 $)
is reproduced.  

Concerning the relative strength of the nucleon and delta vertices, 
we find  that the MCM can give 
a good description of data with soft ($\Lambda \leq 1.0$ $GeV$) cut-off parameters. Choosing 
$\Lambda_{oct} \geq \Lambda_{dec}$ or
$\Lambda_{oct} \leq \Lambda_{dec}$ does not make much difference as long as both 
cut-offs differ only by $\simeq 200$ $MeV$. The latter choice seems to be more appropriate, 
especially if one introduces additional sources of sea asymmetry. 

\vspace{0.5cm}

\underline{Acknowledgements}: This work has been supported by CNPq and  
FAPESP under contract number 98/2249-4. We are indebted to Y. Hama, F.M. Steffens, 
A.W. Thomas and G. Krein for fruitful discussions.

\newpage
\noindent
{\bf Figure Captions}\\
\begin{itemize}

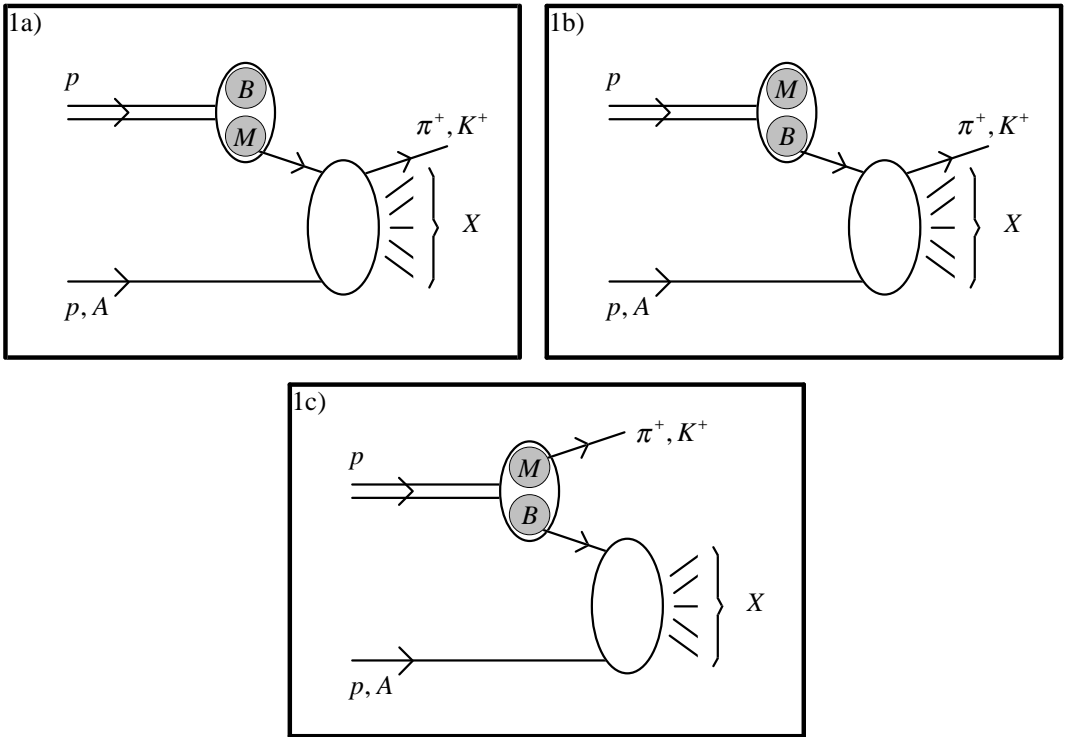
\item[{\bf Fig. 1}] $p p$ or $p A$ collision in which the projectile is in a $|M B>$ state.
a) ``indirect'' $K^+$ or $\pi^+$ production: the meson $M$ from the cloud 
undergoes the reaction $ M \, p (A) \rightarrow \pi^+ (K^+) X$. 
b) ``indirect'' $K^+$ or $\pi^+$ production: the baryon
from the cloud undergoes the reaction $B \, p (A) \rightarrow \pi^+ (K^+) X$. 
c) ``direct'' $K^+$ or $\pi^+$ production: the cloud meson $\pi^+$ ($K^+$) escapes from the
cloud as a spectator.

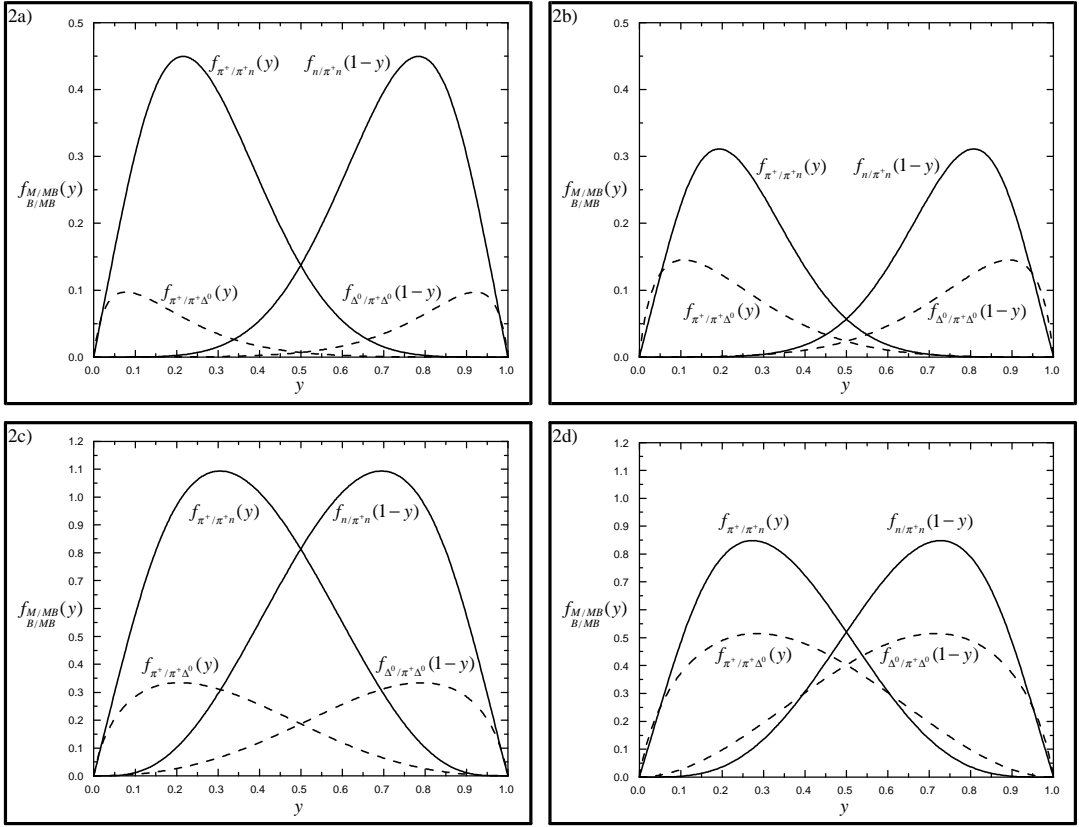
\item[{\bf Fig. 2}] Illustration of the octet (decuplet) splitting functions $f_{M/MB}$ and  
$f_{B/MB}$ in solid (dashed) lines for the following cut-off choices: 
a) $\Lambda_{oct} = 1.0 $ $GeV$ and $\Lambda_{dec} = 0.80 $ $GeV$;  
b) $\Lambda_{oct} = 0.80 $ $GeV$ and $\Lambda_{dec} = 1.0 $ $GeV$;  
c) $\Lambda_{oct} = 2.0 $ $GeV$ and $\Lambda_{dec} = 1.6 $ $GeV$;  
d) $\Lambda_{oct} = 1.6 $ $GeV$ and $\Lambda_{dec} = 2.0 $ $GeV$.

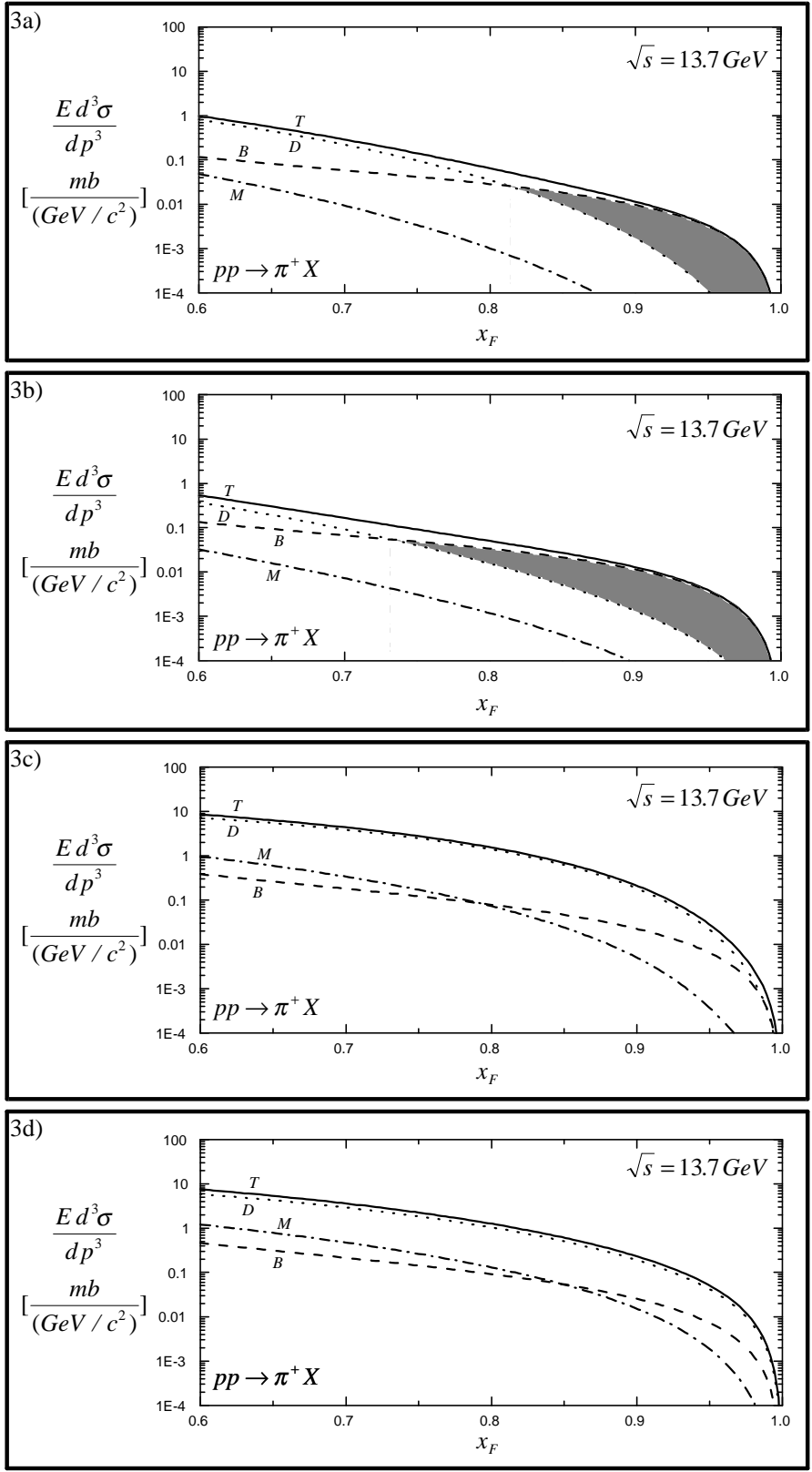
\item[{\bf Fig. 3}] Total ($T$) inclusive pion spectra in $p A$ collisions calculated 
with eq. (\ref{sechoque}) and the direct ($D$), meson initiated ($M$) and  baryon initiated ($B$), 
given respectively by eqs.(\ref{direct}), (\ref{dsigmeson}) and (\ref{dsigbaryon}).
a), b), c) and d) correspond to the cut-off choices made in Fig. 2.

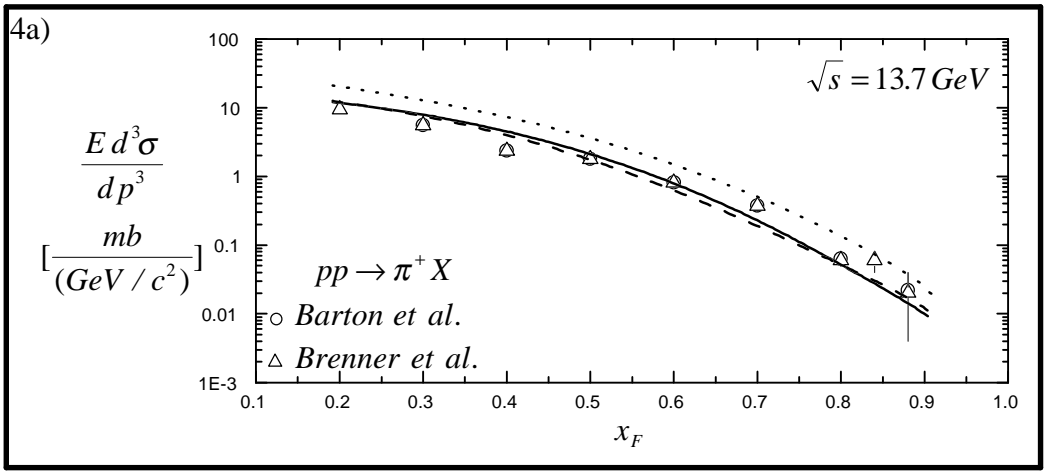
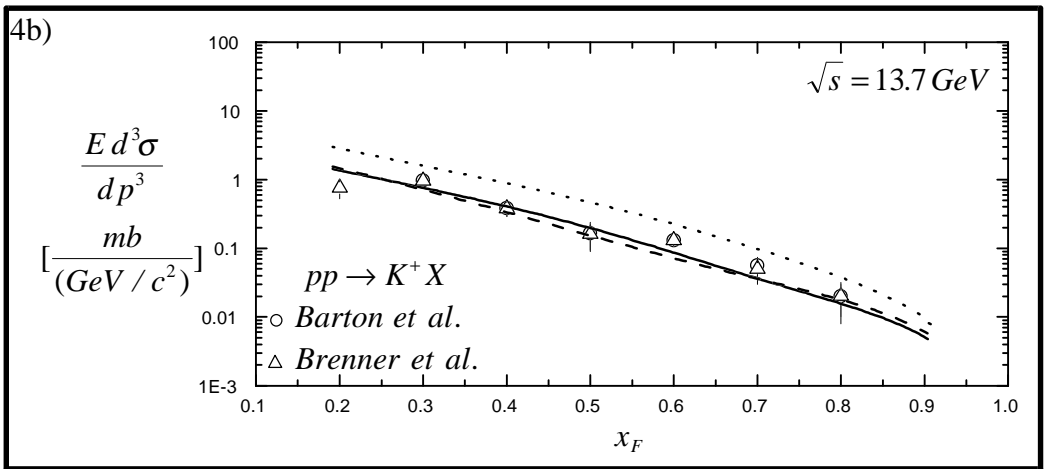
\item[{\bf Fig. 4}] a) Inclusive pion spectra calculated with 
(\ref{sechoque}). Data are from \cite{bar,bren}. Solid, dashed and dotted lines correspond to
the cut-off choices I, II and 
$\Lambda_{\pi N}= 1.0$ $GeV$, $\Lambda_{\pi \Delta}= 1.3$ $GeV$ respectively; b) the same as a) for
kaon spectra.

\item[{\bf Fig. 5}] a) $\overline d - \overline u$ calculated with eq. (\ref{quark}) with 
cut-off choices I (solid line) and II (dashed line) compared with E866 data; b) the same as a)
for the ratio $\overline d / \overline u$.

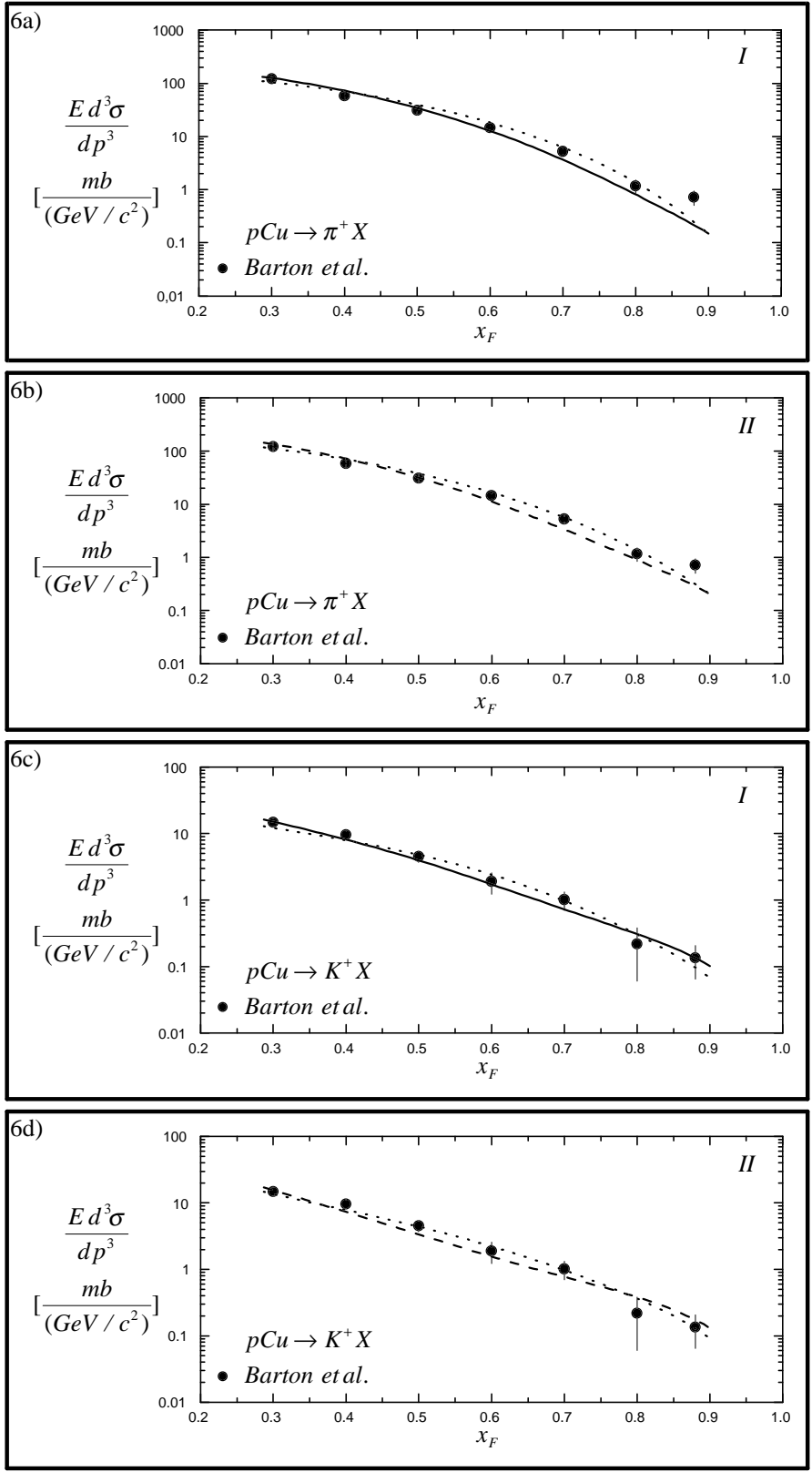
\item[{\bf Fig. 6}] Inclusive meson spectra in proton nucleus collisions calculated with 
(\ref{sechoca}) using cut-off choices I (solid lines) and II (dashed lines). Dotted lines
show the curves obtained taking medium effects into account. Data are from \cite{bar,bren}.
a) pion spectrum with choice I; b) pion spectrum with choice II; c) same as a) for kaons; d) 
same as b) for kaons.

\end{itemize}

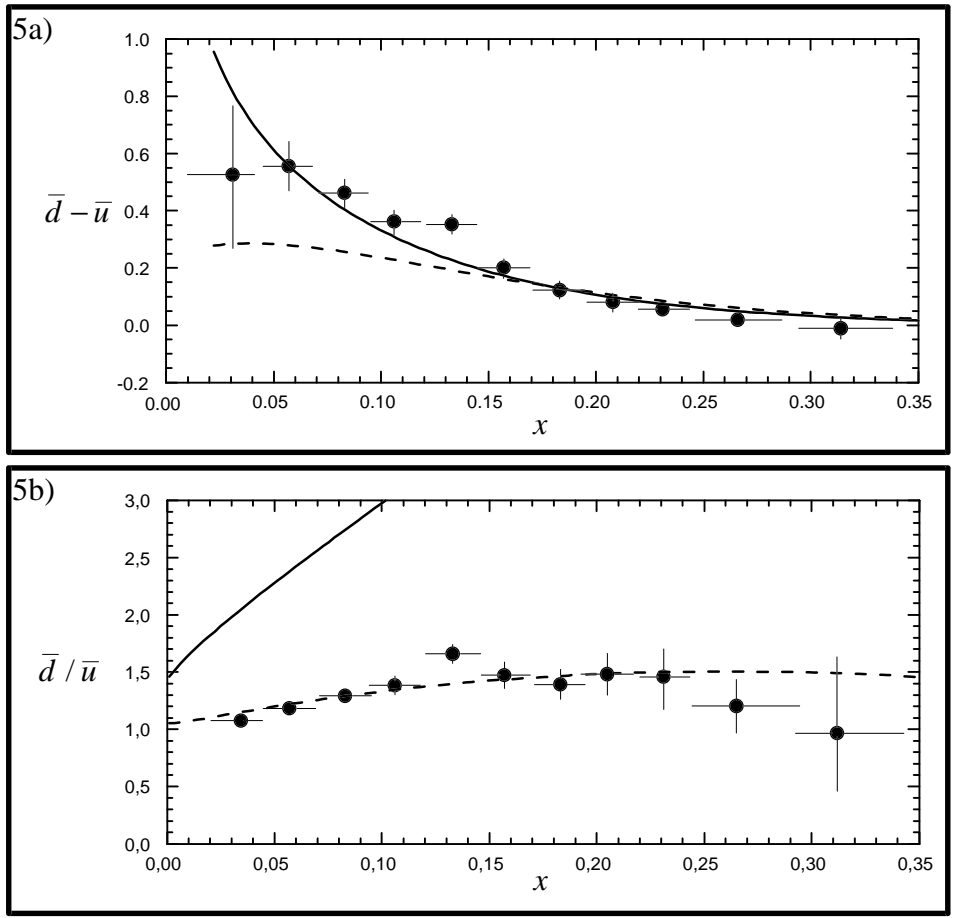
\end{document}